\documentclass[pra,showpacs,twocolumn,amsmath,amssymb,floatfix,superscriptaddress]{revtex4}

\usepackage{graphicx}
\usepackage{pstricks,pst-node,pst-text,pst-3d}
\usepackage{color}

\newcommand{\pdagger}{{\phantom{\dagger}}}
\newcommand{\dt}{\Delta\tau}
\newcommand{\reff}[1]{Fig.\ \ref{fig:#1}}

\newcommand{\myparagraph}[1]{{\it #1} -- }

\newcommand{\neel}{N\'{e}el}

\begin{document}

	\title{N\'{e}el transition of lattice fermions in a harmonic trap: a real-space DMFT study}

\author{E.~V.~Gorelik}
\affiliation{Institute of Physics, Johannes Gutenberg University, 55099 Mainz, Germany}
\author{I.~Titvinidze}
\affiliation{Institut f\"ur Theoretische Physik, Johann Wolfgang Goethe-Universit\"at, 60438 Frankfurt/Main, Germany}
\author{W.~Hofstetter}
\affiliation{Institut f\"ur Theoretische Physik, Johann Wolfgang Goethe-Universit\"at, 60438 Frankfurt/Main, Germany}
\author{M.~Snoek}
\affiliation{Institute for Theoretical Physics, Universiteit van Amsterdam, 1018 XE Amsterdam, The Netherlands}
\author{N.~Bl\"umer}
\affiliation{Institute of Physics, Johannes Gutenberg University, 55099 Mainz, Germany}

\date{\today}

  \begin{abstract}
We study the magnetic ordering transition for a system of harmonically trapped ultracold fermions with repulsive interactions 
in a cubic optical lattice, within a real-space extension of dynamical mean-field theory (DMFT). 
Using a quantum Monte Carlo impurity solver, we establish that antiferromagnetic correlations are
signaled, at strong coupling, by an enhanced double occupancy. This signature is directly
accessible experimentally and should be observable well above the critical temperature
for long-range order.
Dimensional aspects appear less relevant than naively expected.
  \end{abstract}
  \maketitle

Starting with the achievement of Bose Einstein condensation, ultracold atomic gases have led to fascinating insights in quantum many-body phenomena \cite{Bloch08_RMP}. With the recent experimental successes in realizing quantum degenerate \cite{Koehl05} and strongly interacting \cite{Esslinger08_Nature,Bloch08_ferm} Fermi gases in optical lattices, such systems are widely considered as highly tunable {\it quantum simulators} of condensed matter \cite{Bloch08_Science}: 
in principle, most solid-state Hamiltonians of interest are accessible by selecting appropriate fermionic 
flavors, interactions, and lattice parameters \cite{Zoller05}. 
The recent observation of the fermionic Mott transition in binary mixtures of $^{40}K$ in three-dimensional simple-cubic lattices \cite{Esslinger08_Nature,Bloch08_ferm} marks important progress in this respect.

A major current challenge is the realization and detection of quantum magnetism, 
i.e. spontaneous magnetic order, in ultracold atoms. 
For bosons in optical lattices, 
correlated particle tunneling \cite{Foelling07} 
and superexchange \cite{Trotzky08} -- basic mechanisms underlying quantum magnetism -- have already been observed. 
Moreover, indications for a ferromagnetic Stoner instability have recently been found in a trapped spin-1/2 Fermi gas \cite{Jo09}.  
However, the antiferromagnetic (AF) phase, which is generic for Hubbard-type models and many classes of solids at low temperatures, has not yet been seen in cold fermions in optical lattices. 
So far, experimental progress in this direction is mainly sought by trying to cool the systems
down to a central entropy per particle $s\lesssim \log(2)/2$ required for long-range AF order on a cubic lattice \cite{De_Leo_PRL08,Wessel10}
(which is a factor of 2 below current average entropies \cite{Joerdens09}).

Due to the intrinsic inhomogeneity of trapped atomic clouds, even the qualitative interpretation of experimental data 
may rely on corresponding quantitative simulations (e.g. of cloud compressibilities across the Mott transition \cite{Bloch08_ferm}).
An appropriate quantitative theory of AF ordering in 
fermions in an optical lattice
should therefore capture strong correlation effects as well as the spatial inhomogeneity 
at the temperatures of experimental interest. 
The dynamical mean-field theory (DMFT) is well established 
as a powerful, nonperturbative approach to interacting Fermi systems in three dimensions 
\cite{Georges96,Kotliar_Vollhardt}. 
 Spatial inhomogeneities of the optical lattice can be captured within a real-space 
extension of the method (RDMFT) \cite{Snoek_NJP08,Helmes_PRL08} 
or, approximately, by applying DMFT within a local density approximation (LDA). 
In either case, the accessible parameter ranges and the numerical accuracy depend on the method 
chosen as DMFT impurity solver.
So far mainly the numerical renormalization group (NRG) has been used in this context, which 
is reliable at low temperature $T$, but leads to artifacts at elevated $T$ \cite{Bloch08_ferm}. 
Thus, NRG is only adequate for studying effects with weak temperature
dependence, such as the paramagnetic Mott transition \cite{Rosch08PRL,Bloch08_ferm}, or for
ground state investigations
\cite{Snoek_NJP08}. In contrast, quantum Monte Carlo (QMC) based methods are precise
and even numerically cheap at or above the $T$ ranges relevant for AF ordering and accessible
in cold-atom experiments.

Observables routinely measured in solids, e.g., using x-ray or neutron diffraction, 
may be hard to access in cold-atom based systems. In particular, the detection of
the AF order parameter, although in principle possible via noise correlations \cite{Altman04} or Bragg spectroscopy \cite{Corcovilos10},
is highly nontrivial. It is, therefore, essential to identify fingerprints of AF phases that are 
easily accessible in current experiments. Ideally, such observables should also be sensitive to
precursor effects, monitoring the approach to the ordered phase.

In this Letter, we present Hirsch-Fye QMC \cite{Hirsch86} based RDMFT studies of 
trapped spin-$1/2$ lattice fermions at low to intermediate $T$,
employing a massively parallel code \cite{Bluemer10CCP}. 
We demonstrate that the onset of AF correlations at low $T$ is signaled, for 
sufficiently strong interactions, by a significantly enhanced double occupancy $D$. 
This signature, which had been missed so far even in studies of the homogeneous Hubbard model,
appears ideal in the cold-atom context, since (i) $D$ is directly accessible 
experimentally and (ii) its enhancement measures nearest-neighbor AF correlations
which are expected as a precursor effect already above the \neel\ temperature.
We find strong proximity effects (beyond LDA) at the interfaces between AF
and paramagnetic regions, implying that RDMFT is important for
quantitative studies of ordering phenomena in inhomogeneous systems. 
Our detailed predictions provide essential guidance to experimentalists.


\myparagraph{Model and Methods}
Balanced mixtures of spin-$1/2$ fermions in an optical lattice are well described by the Hubbard
model (with trapping potential $V_i = V_0 r_i^2/a^2$),

\vspace{-3ex}
  \begin{equation}\label{Hubb_mod}
    \hat{H} =\! -t \sum_{\langle ij\rangle ,\sigma}  \hat{c}^{\dag}_{i\sigma}
  \hat{c}^\pdagger_{j\sigma} 
  +  U \sum_{i} \hat{n}_{i\uparrow} \hat{n}_{i\downarrow}
  \,+ \sum_{i, \sigma}(V_i - \mu)\, \hat{n}_{i\sigma}.
  \end{equation}

	\vspace{-1ex}
\noindent
  Here, $\hat{n}_{i\sigma} = \hat{c}^{\dag}_{i\sigma} \hat{c}^\pdagger_{i\sigma}$, 
$\hat{c}^\pdagger_{i\sigma}$ ($\hat{c}^{\dag}_{i\sigma}$) are annihilation (creation)
operators for a fermion with (pseudo) spin $\sigma\in \{\uparrow,\downarrow\}$ at site $i$ (with coordinates ${\bf r}_i$), 
$t$ is the hopping amplitude between nearest-neighbor sites $\langle ij\rangle$, $U > 0$
is the on-site interaction, and $\mu$ is the chemical potential.
In the following, we choose $V_0=0.05t$ for a nearly realistic confinement \cite{fn:trap}
and use $t$ as the energy unit.

The RDMFT approach \cite{Snoek_NJP08,Helmes_PRL08} is based on the following
expression for the Green function matrix $G$ for spin $\sigma$ and Matsubara frequency
$i\omega_n=i(2n+1)\pi T$: 

\vspace{-3ex}
\begin{equation}
\big[{G_\sigma}({i\omega_n})\big]^{-1}_{ij} \! = t_{ij} + \big[{i\omega_n} + {\mu} - {V_i} - {\Sigma_{i\sigma}}({i\omega_n})\big]\, \delta_{ij}\,. 
\end{equation}

\vspace{-1ex}\noindent
Here the only approximation is the DMFT assumption of a local (i.e. site-diagonal) self-energy $\Sigma$,
which corresponds to the momentum independence of $\Sigma$ in translation-invariant systems \cite{Kotliar_Vollhardt}.
The standard DMFT impurity problems, 
one for each inequivalent lattice site, are solved using the Hirsch-Fye QMC algorithm \cite{Hirsch86}; Trotter errors in the raw data 
are eliminated by extrapolation of the imaginary-time discretization $\dt\to 0$ \cite{Bluemer05,Bluemer07}. Due to the relatively
high $T$ considered here, statistical errors can be avoided as well in full enumerations of the Hubbard-Stratonovich
fields, which improves the data for ordered phases at large $U\gtrsim 12t$. The resulting precision is
much better than achievable with NRG \cite{Bluemer10CCP}. This remains true within the ``slab'' approximation, in which the full system is reconstructed from the properties of a central set of planes with periodic boundary conditions in all directions \cite{Bluemer10CCP}. 
The entropy $S$ is not directly accessible in QMC simulations, therefore we use the
thermodynamic relation 
$S(\mu,T) = \int_{-\infty}^\mu d\mu' \,\partial N(\mu',T) / \partial T$ (for total particle number $N$) \cite{Bloch08_ferm,Bluemer10CCP}.

\vspace{0.5ex}
\myparagraph{Results}
We consider a fermionic cloud confined in a cubic lattice for $\mu=U/2$, i.e. for half filling at the center (at least within LDA) and
focus, initially, on a moderatly strong interaction $U=12t$, equal to the bandwidth $W$. 
In \reff{melting},
\begin{figure} 
	\includegraphics[width=\columnwidth]{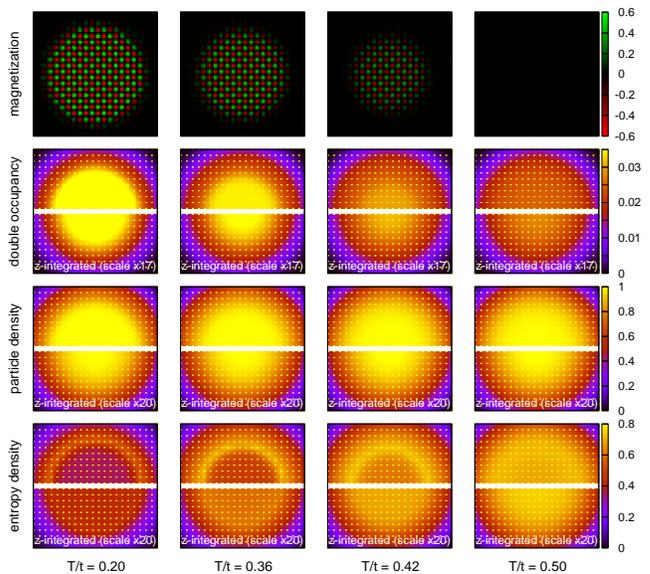}
\caption{(Color online) First row: AF order in central plane of 
atomic cloud (in cubic optical lattice) for intermediate coupling $U=W=12t$. At low
$T\lesssim 0.2t$ (left column), a large AF core is strongly polarized (nearly 90\%);
with increasing $T$, both extent and magnitude of the AF order decay.
Second row: the double occupancy (upper/lower half: central/column density)
is strongly enhanced in AF regions; in contrast, the density ($3^{\text{rd}}$ row, 
central/column) shows little $T$ dependence. 
$4^{\text{th}}$ row: entropy densities (central/column).}
\label{fig:melting} 
\end{figure}
the core shows a nearly perfect staggered magnetization (first row) at low $T$ (left column). With increasing $T$ (from left to right), both the extent of the ordered region and its polarization decrease until the order is lost at the bulk N\'{e}el temperature $T_{\text{N}} \approx 0.46t$. 
Unfortunately, this most obvious signature of AF order is not directly accessible experimentally, 
since single-site resolution \cite{Greiner09_Nature} has not yet been achieved for three-dimensional systems.
The \neel\;transition is also not detectable via the particle density profiles (third row), as they hardly change at this scale. In contrast, the double occupancy $D_i=\langle n_{i\uparrow}n_{i\downarrow}\rangle$
provides a pronounced signal: at high $T$, it is featureless in the center, with a maximum value of about $0.025$. Only at low $T$, it is enhanced, by up to $50\%$, in the emerging central antiferromagnetic core. 
This signal remains strong even after integrating over the line of sight ($z$ axis) as in most experimental detection schemes; resulting column density plots are shown in the lower halves of the split images in \reff{melting}.
We note that total double occupancies 
have already been measured with relative accuracies of $20\%$ \cite{Esslinger08_Nature}; a higher precision and spatial resolution appear feasible. 
Entropy, on the other hand, cannot be measured {\em in situ}; thus, its suppression in the AF core (\reff{melting} bottom) is not visible experimentally.

The observed enhancement of $D$ is also present in the homogeneous Hubbard model at half filling (cf. \reff{AF_scheme_half-filling}), for which the underlying mechanism is most easily explained. For high $T$ {\em thermal fluctuations} prevail, which increase $D$ with increasing temperature towards the uncorrelated limit $D_i\to \langle n_{i\uparrow}\rangle\langle n_{i\downarrow}\rangle$; this effect has already been used in thermometry (comparing, e.g., to high-$T$ expansions) \cite{Joerdens09}. Conversely, {\em quantum fluctuations} in the metallic Fermi-liquid regime can significantly increase $D$ for $T\to 0$; this manifestation of the Kondo effect has been discussed in the context of Pomeranchuk cooling (without taking the low-$T$ AF phase into account) \cite{FWerner05,Dare07,De_Leo_PRL08}.
In contrast, the enhancement seen in \reff{melting} is a {\em strong-coupling effect} close to half filling
as illustrated in \reff{AF_scheme_half-filling}:
\begin{figure}[t]
\unitlength0.1\columnwidth
\begin{picture}(10,4.5)
\put(0,0){\includegraphics[width=0.45\columnwidth]{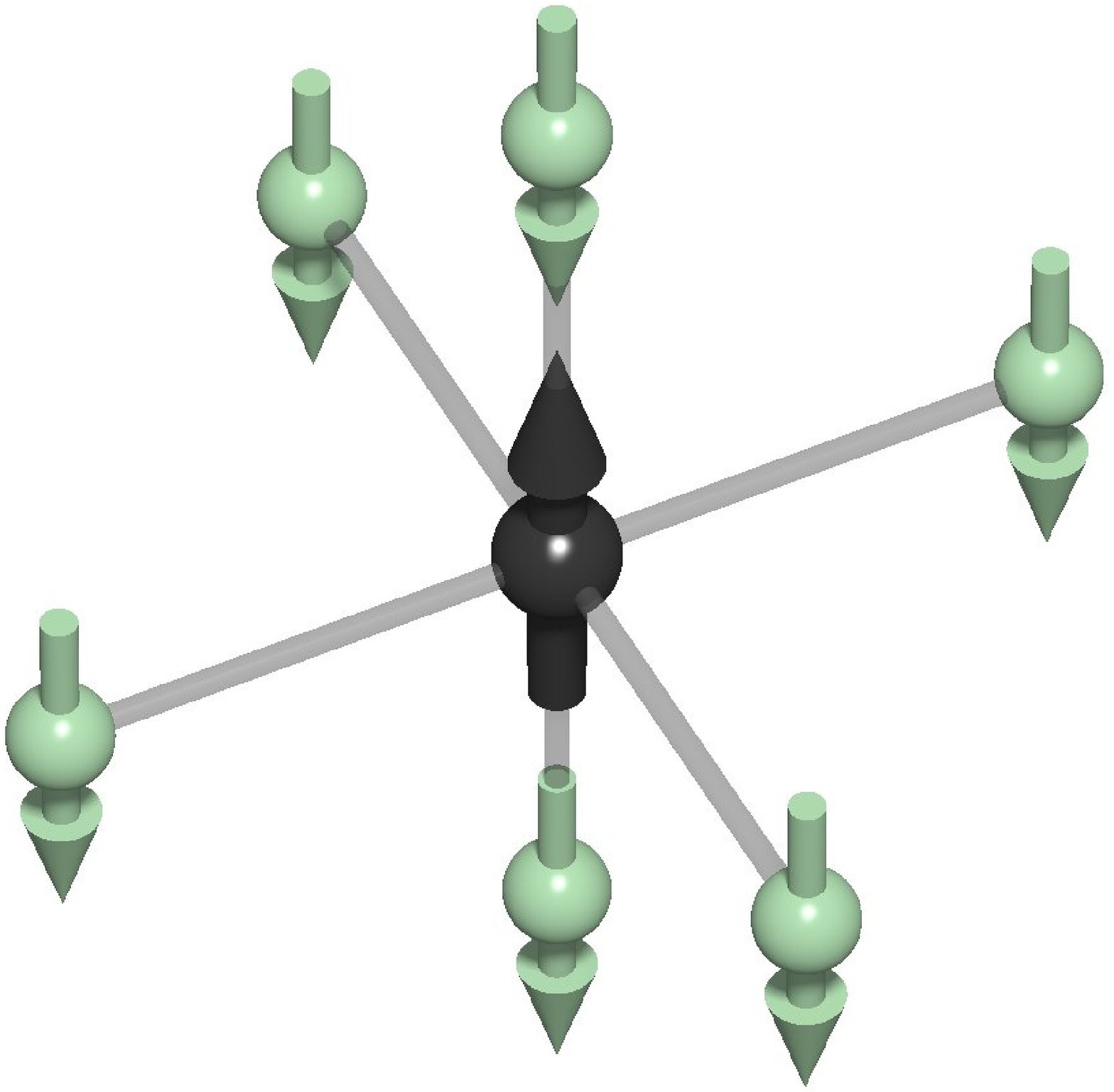}}
\put(5.5,0){\includegraphics[width=0.45\columnwidth]{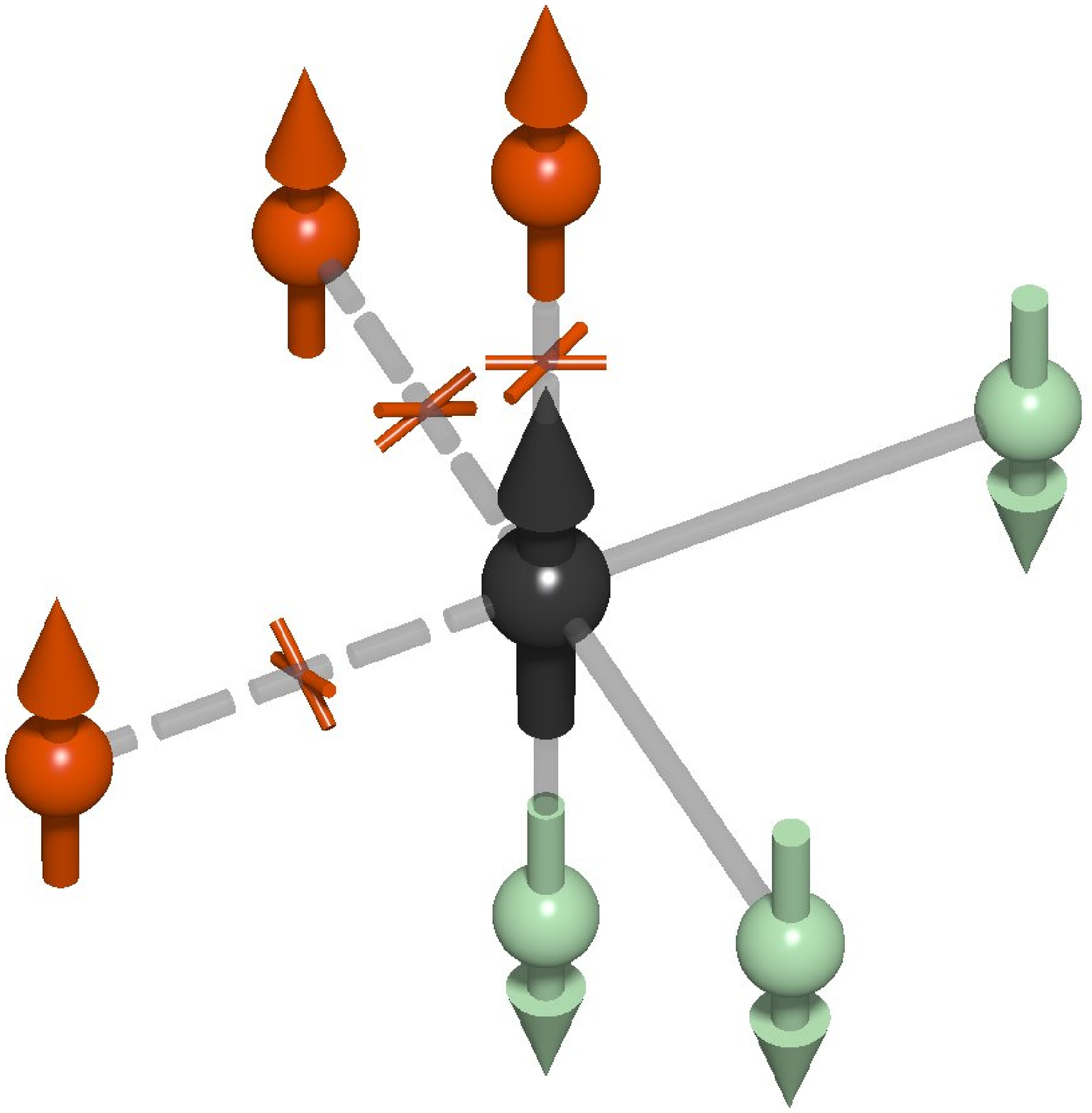}}
\put(0,4){(a)}
\put(5.5,4){(b)}
\put(2.5,2.1){\includegraphics[width=0.125\columnwidth]{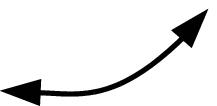}}
\put(3.1,1.8){$\displaystyle -\frac{t^2}{U}$}
\put(8,2.1){\includegraphics[width=0.125\columnwidth]{arrow}}
\put(8.6,1.8){$\displaystyle -\frac{t^2}{U}$}
\end{picture}

	\includegraphics[width=\columnwidth]{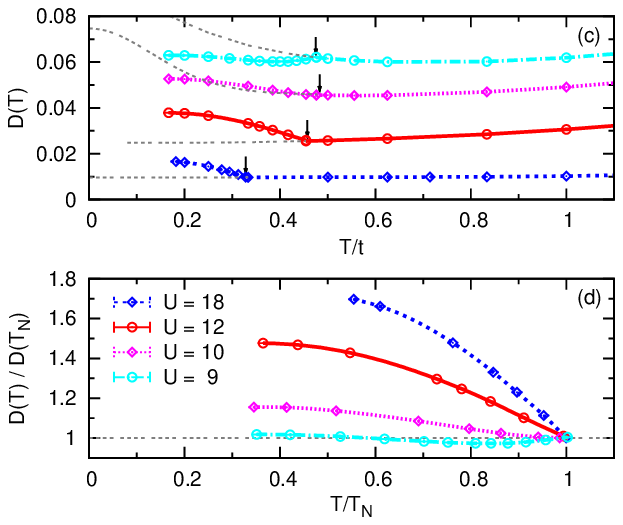}
\caption{(Color online) 
Illustration of mechanism for enhanced double occupancy (at strong coupling) 
in the AF state (a) compared to a paramagnetic state (b); see text.
(c) DMFT-QMC estimates of double occupancy 
at half filling versus $T$ for various interactions $U$; arrows indicate
corresponding N\'{e}el temperatures. Thin lines: results for metastable 
paramagnetic phase. (d) same data scaled to values of critical point.
}\label{fig:AF_scheme_half-filling} 
\end{figure}
In a fully developed AF state (a), a central spin-up atom (central black arrow) can hop virtually to all $Z=6$ next neighbors, lowering its energy (in $2^{\text{nd}}$ order) to $E_{\text{AF}}=-Z t^2/U$.
In contrast, half of the neighboring sites are (on average) forbidden by the Pauli principle in a paramagnetic state (b),
thus $E_{\text{p}}=-Z t^2/(2U)$. 
By $D=dE/dU$ (valid at $T=0$), the argument implies $D_{\text{AF}}/D_{\text{p}}\stackrel{U\to\infty}{\longrightarrow}2$. 
Thus, nearest-neighbor AF correlations can double $D$ compared to the paramagnet at large $U$ and low $T$.  Ferromagnetic correlations, on the other hand, would suppress $D$.

The DMFT data shown in \reff{AF_scheme_half-filling} (c) for a half-filled homogeneous system confirm this (dimension independent) picture: At $U=18t$ (dashed line), the double occupancy increases strongly below $T_N\approx 0.33t$ (arrow); the relative enhancement at $T\approx 0.2t$ exceeds $70\%$, as best seen in the scaled view of \reff{AF_scheme_half-filling} (d). In contrast, no temperature dependence is visible in the paramagnetic phase, which may be continued artificially 
(thin lines) down to $T=0$. Although the relative AF enhancement of $D$ is smaller at $U=12t$ (solid lines), both the larger absolute scale and the larger $T_{\text{N}}\approx 0.46$ make this parameter more favorable for experiment; 
whereas thermal fluctuations are still small at $T\approx T_{\text{N}}$. The strong-coupling effect vanishes rapidly for smaller $U$; at $U=9t$ (dash-dotted line), the AF correlations even lead to a reduction of $D$.

\begin{figure}[t]
	\includegraphics[width=\columnwidth]{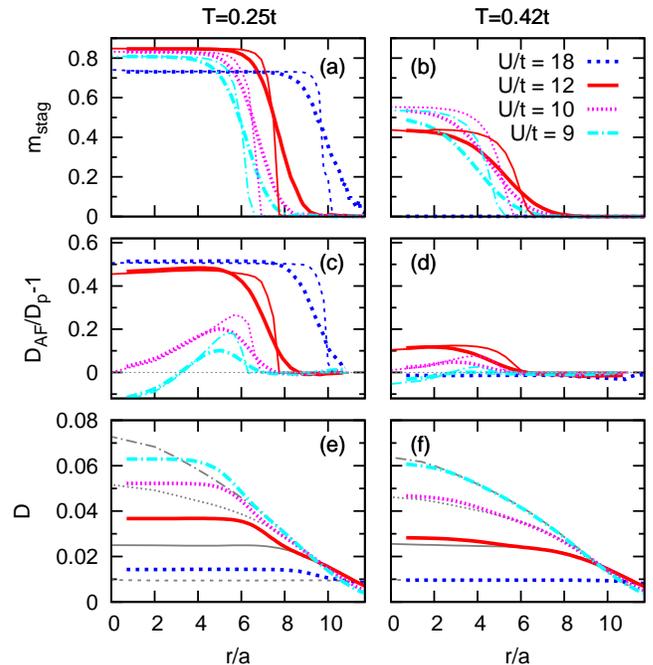}
\caption{(Color online) 
Radial dependence of staggered magnetization (top row) and double 
occupancy (bottom row) for $V=0.05t$ at temperature $T=0.25t$ (left column) and $T=0.42t$ (right column).
The central row shows the relative enhancement of $D$ in the AF phase.
Thick lines correspond to RDMFT data; thin lines represent LDA results in (a)-(d)
and paramagnetic calculations in (e)-(f).
}\label{fig:mstag_dop_vs_r} 
\end{figure}
For quantitative predictions, let us now return to RDMFT results for inhomogeneous systems.
As seen in \reff{mstag_dop_vs_r} (a), the order parameter (thick lines) is nearly constant in the AF core at $T=0.25t$, with $m_{\text{stag}}^{\text{max}}\gtrsim 0.8$ for $U/t=9,10,12$ and  $m_{\text{stag}}^{\text{max}}\approx 0.7$ for $U=18t$, before it slowly decays to zero. The width of the transition region (3 to 4 lattice spacings) must be attributed to proximity effects \cite{Snoek_NJP08} since the phase boundaries are sharp within LDA (thin lines). 
The double occupancies (thick lines in panel e) show strong enhancement throughout the AF core in comparison with (enforced) paramagnetic solutions (thin lines) for $U/t=12,18$; for $U\lesssim 10t$, the latter yield even larger $D$ in the center. As seen in \reff{mstag_dop_vs_r} (c), the relative enhancement of $D$ is almost equal for $U=12t$ and $U=18t$ at this fixed temperature (cf. \reff{AF_scheme_half-filling}c); the strong deviations between RDMFT and LDA solutions indicate significant proximity effects for all $U$.
At larger $T=0.42t$, the AF order is generally weakened (and disappears for $U=18t$) as shown in \reff{mstag_dop_vs_r} (b); the transitions also become smoother, especially towards the shrinking core. Both effects result in a smaller enhancement of D for $U=12t$ (\reff{mstag_dop_vs_r}(d) and (f)).

Let us, finally, discuss whether the double occupancy retains its clear signals for
AF correlations in averages over large parts of the cloud \cite{Esslinger08_Nature}.
This is indeed the case, as shown in \reff{Dfrac_integrated} for $U=12t$: although the
total fraction $D_{\text{frac}}=2\sum_i D_i/N$ of atoms on doubly occupied sites (circles) is temperature
dependent also at elevated temperatures, due to the impact of the metallic shell, the
strong increase in the AF phase at $(T/t)^2\lesssim 0.2$
can clearly be distinguished from the extrapolated paramagnetic behavior (thin dotted line).
As expected, the signal becomes sharper when the measurements are concentrated \cite{fn:gauss}
towards the AF core;  
for $R=2 a$ (triangles in \reff{Dfrac_integrated}) the slope of the curve is negligible above $T_{\text{N}}$, 
so that the increase of $D_{\text{frac}}$ with the onset of AF is almost as pronounced 
as in the homogeneous case (\reff{AF_scheme_half-filling}).
\begin{figure}[t]
\unitlength0.1\columnwidth
\begin{picture}(10,6.4)
\put(0,-0.5){\includegraphics[width=\columnwidth]{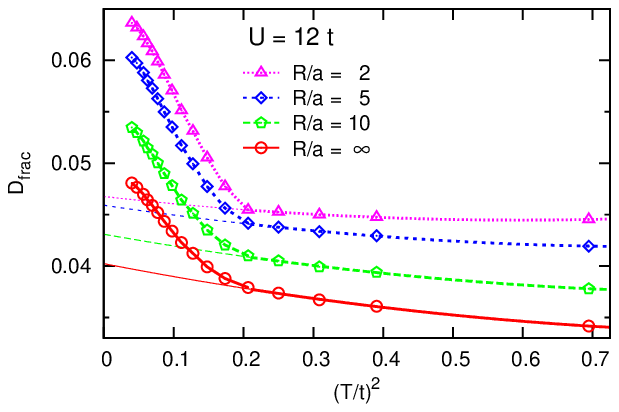}}
\put(6.8,3){\includegraphics[width=0.27\columnwidth]{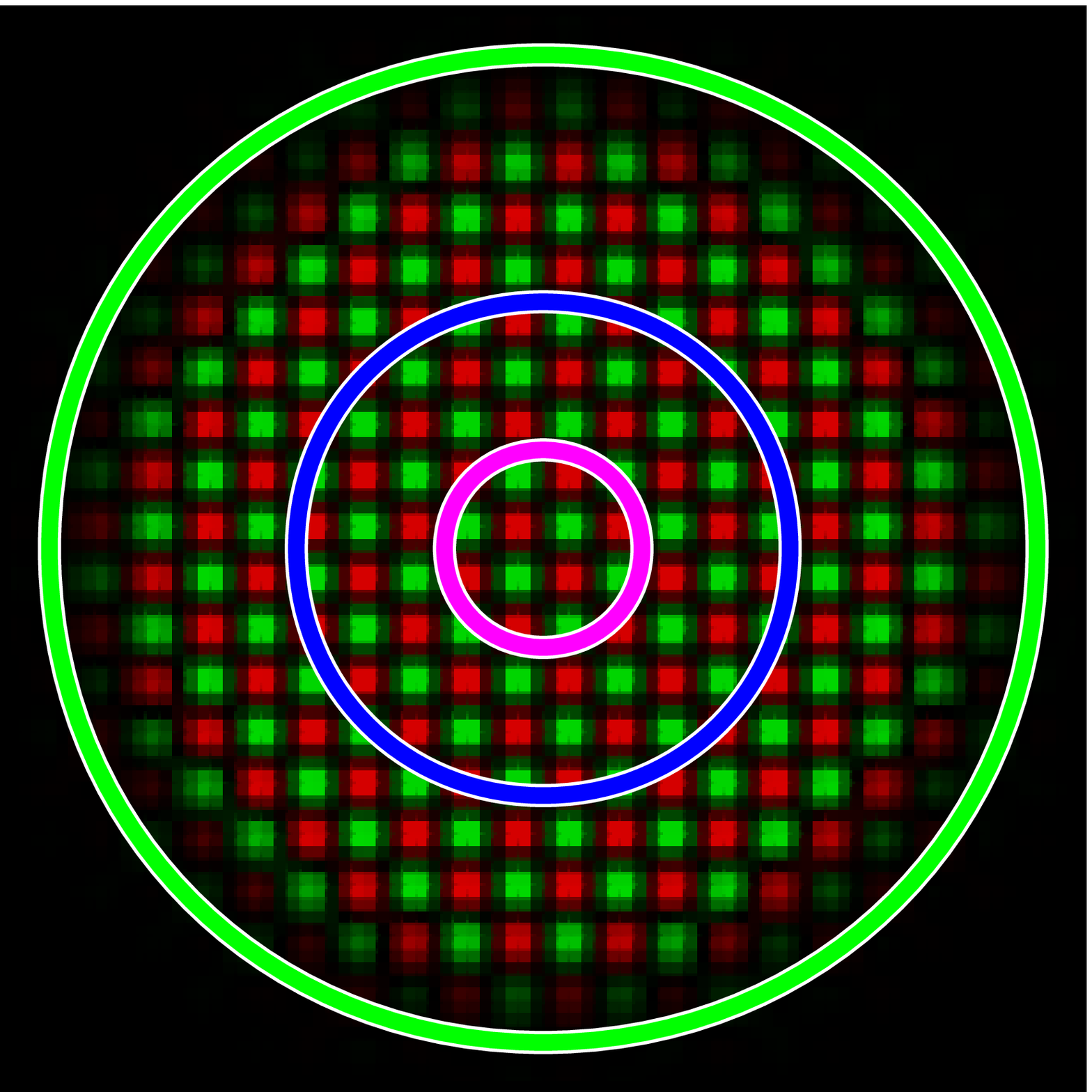}}
\end{picture}
\caption{(Color online)
Fraction of atoms on doubly occupied lattice sites as a function of squared temperature;
circles represent RDMFT results for the full cloud; polygons correspond to measurements
along a gaussian beam of radius $R$, as illustrated in the inset.
Thick lines are quadratic fits for $(T/t)^2\ge (T_{\text{N}}/t)^2\approx 0.21$ (and guides to
the eye below); their extrapolations $T\to 0$ are shown as thin lines.
}\label{fig:Dfrac_integrated} 
\end{figure}

\myparagraph{Conclusion} 
In this Letter, we have presented the first simulation of the \neel\ transition in fermions confined in an optical lattice at full DMFT accuracy, using a new QMC based RDMFT implementation for realistic trap parameters. We have found that the onset of AF order at low $T$ is signaled in the strong-coupling regime ($U\gtrsim W=12t$) by an enhanced double occupancy. This is a very specific signature of antiferromagnetism since $D$ is (nearly) temperature independent in the paramagnetic phase in this regime (for $T\lesssim t/2$ and within DMFT). 

It is clear that nonlocal correlations (beyond DMFT) will affect some of the results of this study: e.g., {\em long-range} AF order can be expected in 3 dimensions only about $25\%$ below $T_{\text{N}}^{\text{DMFT}}$ \cite{Staudt00}. On the other hand, DMFT can reproduce recent QMC measurements of $D(T)$ in 2 dimensions \cite{Paiva10} (for $U/t=8$ \cite{fn:scaling} and low $T$) within $10\%$; even the temperature scale $T_{\text{spin}}$ for {\em short-range AF correlations} \cite{Paiva10} is only $10\%$ below $T_{\text{N}}^{\text{DMFT}}$ (at $U/t=8$). We may therefore expect our predictions for $D(T)$ to be quantitatively accurate in 3 dimensions, up to a shift of the kink from $T_{\text{N}}^{\text{DMFT}}$ to the true $T_{\text{N}}$. In particular, the low-temperature enhancements of $D$ should set in around $T_{\text{N}}^{\text{DMFT}}$, as a signature of increasing {\em short-range AF correlations}. High-precision measurements of the double occupancy will thus be able to reveal AF correlations (as a precursor for AF order) at essentially the temperatures/entropies reached in current experiments.

  We thank I.\ Bloch, P.G.J.\ van Dongen, A.\ Georges, H.\ Moritz, U.\ Schneider, and S.\ Wessel for valuable discussions.
  Support by the DFG within the Sonderforschungsbereich SFB/TR 49 and by the John von Neumann Institute for Computing is
  gratefully acknowledged.


\begin{thebibliography}{16}
\bibitem{fn:gauss} We assume a detection probability for density and double occupation at site $\vec{r}_i$ of $p=\exp[-(x_i^2+y_i^2)/R^2]$.
\end{thebibliography}
\end{document}